\documentclass[twocolumn,aps,prb,showpacs]{revtex4-1}

\usepackage{amsmath}
\usepackage{amssymb}
\usepackage{upgreek}
\usepackage{float}
\usepackage{graphicx}
\usepackage{wrapfig}
\usepackage{natbib}
\usepackage{color}
\usepackage{epstopdf}
\usepackage[singlelinecheck=off, justification=raggedright]{caption}
\usepackage{subcaption}

\usepackage{etex}
\reserveinserts{32}
\usepackage{morefloats}

\newcommand{\vect}[1]{{\boldsymbol{#1}}}

\makeatletter

\begin{document}

\title{Dielectric screening of surface states in a topological insulator.}

\author{J. P. F. LeBlanc$^{1}$}
\email{jpfleblanc@gmail.com}
\author{J. P. Carbotte$^{2,3}$}

\affiliation{$^1$Max-Planck-Institute for the Physics of Complex Systems, 01187 Dresden, Germany} 
\affiliation{$^2$Department of Physics and Astronomy, McMaster
University, Hamilton, Ontario L8S~4M1 Canada}
\affiliation{$^3$The Canadian Institute for Advanced Research, Toronto, ON M5G~1Z8 Canada}


\date{\today}
\begin{abstract}
Hexagonal warping provides an anisotropy to the dispersion curves of the helical Dirac fermions that exist at the surface of a topological insulator.  A sub-dominant quadratic in momentum term leads to an asymmetry between conduction and valence band.  A gap can also be opened through magnetic doping.  We show how these various modifications to the Dirac spectrum change the polarization function of the surface states and employ our results to discuss their effect on the plasmons.  In the long wavelength limit, the plasmon dispersion retains its square root dependence on its momentum, $\vect{q}$, but its slope is modified and it can acquire a weak dependence on the direction of $\vect{q}$.  Further, we find the existence of several plasmon branches, one which is damped for all values of $\vect{q}$, and extract the plasmon scattering rate for a representative case.
\end{abstract}

\maketitle

\section{Introduction}\label{sec:intro}
The dielectric properties of Dirac fermions have been extensively studied  since graphene, a single monolayer of carbon atoms, was first isolated.\cite{wunsch:2006, dassarma:li:2013:2, hwang:2007, principi:2009, neto:2009, kotov:2012, novoselov:2004}  The electron dynamics in this two dimensional membrane, remarkably, are governed by the relativistic Dirac equation and this has many consequences such as a distinctive signature in the integer quantum Hall effect.\cite{zhang:2005, novoselov:2005, gusynin:2005}  More recent works on the density-density correlation function include extensions to account for a mass term\cite{scholz:2011, scholz:2013} and both Rashba and intrinsic spin-orbit coupling.\cite{scholz:2012}  These are relevant to topological insulators which are insulating in the bulk with metallic surface states protected by topology which exhibit a Dirac spectrum between bulk bands.\cite{hasan:2010, qi:2011, moore:2010, hsieh:2009, chen:2009} While in graphene, the Dirac charge carriers have a pseudospin associated with the two atoms per unit cell honeycomb lattice in topological insulators, the spins are real electron spins with spin-momentum locking.\cite{hsieh:2009}
Unlike graphene, a gap in the energy spectrum of the helical Dirac electrons can be opened by doping with magnetic impurities.\cite{chen:2010}
While intrinsic graphene is often described by models with particle-hole symmetry, topological insulators show asymmetry, modelled by an additional quadratic\cite{wright:2013, taskin:2011, li:2013:2} in momentum (Schr\"odinger) term  in their energy dispersion curves in addition to the dominant linear Dirac term.  This leads to a goblet or hourglass shape \cite{hsieh:2009, xu:2011, qi:2010} which replaces the perfect Dirac cones of graphene with a surface state valence band which fans out in relation to the surface state conduction band.
There is also an important hexagonal warping contribution \cite{fu:2009, li:2013, xiao:2013} to the surface state Hamiltonian.  This leads to significant changes in the associated Fermi surface which starts as circular for small values of chemical potential, $\mu$, and gradually acquires a hexagonal or snowflake shape as $\mu$ is increased.  
This change in geometry has been observed in angular resolved photoemission (ARPES) data.\cite{chen:2009}  Fu\cite{fu:2009} showed that the Fermi surface data could be understood by adding a hexagonal warping cubic term to the Hamiltonian and it has been subsequently shown\cite{li:2013} that this term can have a profound effect on interband optical transitions.  While in graphene the interband transitions lead to a constant uniform background conductivity \cite{gusynin:2007, gusynin:2009, li:2008} of $\sigma_0 = \pi e^2/2h$, the inclusion of hexagonal warping leads instead to a background which increases with increasing photon energy above the threshold for interband absorption which has an onset at twice the value of the chemical potential.\cite{li:2013}

\begin{figure}
  \begin{center}
  \includegraphics[width=0.95\linewidth]{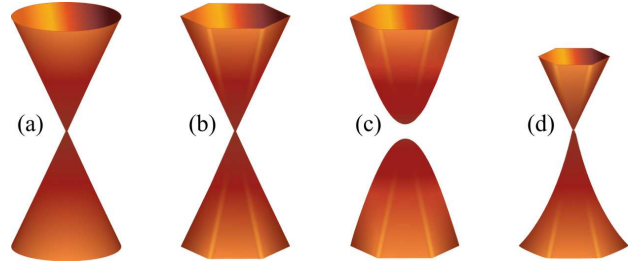}
  \end{center}
 \caption{\label{fig:hex}(Color online) Schematic energy dispersions for (a) only the Dirac term, (b) both Dirac and hexagonal warping terms, (c)Dirac, hexagonal warping and gap. (d) Dirac, hexagonal warping and Schr\"odinger quadratic in momentum term. }
\end{figure}

In this paper we collect these contributions, hexagonal warping, a gap and a sub-dominant quadratic in momentum term and study their effects on the dielectric screening properties of the surface carriers.\cite{bianchi:2010:nc,hwang:2009, adam:2012, dassarma:li:2013}  The hexagonal warping is particularly interesting because it leads to a directional anisotropy.   
The density-density response, or polarization function, $\Pi(\vect{q},\omega)$, can then depend on the angle of the scattering momentum vector $\vect{q}$ defined relative to the $\Gamma \to K$ direction in the  hexagonal honeycomb lattice.  This anisotropy is expected to grow as the chemical potential is increased and the shape of the Fermi surface begins to deviate more from circles.  Consequently, the plasmons which form in the system will depend not only on the absolute value of their momentum, but will also depend on angle.
Recently, Di~Pietro et al.\cite{dipietro:2013} have reported experimental results of Dirac plasmons in the topological insulator Bi$_2$Se$_3$ but did not consider warping effects in their analysis.  This motivates us to fully study what effect, if any, the warping has on the plasmon dispersion.

In section \ref{sec:mod} we specify our model Hamiltonian and give the expression for the polarization function, $\Pi(\vect{q},\omega)$, as a function of scattering momentum, $\vect{q}$, and energy, $\omega$.  Numerical results for the real and imaginary parts of $\Pi(\vect{q},\omega)$ are presented in Sec.~\ref{sec:num}.  We also provide color map plots for the imaginary part of the inverse dielectric function.  Section \ref{sec:anal} deals with the plasmon dispersion, $\omega_p(\vect{q})$, wherein we also provide simplified expressions for the slope of $\omega_p(\vect{q})$ in the long wavelength limit.  Numerical results for the plasmon dispersion are also provided which go beyond the small $q$ limit.  A summary of our findings and concluding remarks are found in Sec.~\ref{sec:conc} followed by a brief appendix which contains relevant algebra.
  
\section{Model and Polarization}\label{sec:mod}

We begin with the Kane Mele Hamiltonian\cite{kane:2005} for helical Dirac fermions at the $\Gamma$ point of the surface state Brillouin zone of a topological insulator which further includes a gap, $\Delta$, a cubic hexagonal warping term of strength $\lambda$, and a sub-dominant quadratic in momentum Schr\"odinger term.  Together these can be written as
\begin{equation}\label{eqn:H}
H=\hbar v(k_x \sigma_y -k _y \sigma_x) + \frac{\lambda}{2}(k_+^3 + k_-^3) \sigma_z +\Delta \sigma_z +E(k)
\end{equation}
where $\sigma_x$, $\sigma_y$, and $\sigma_z$ are the Pauli spin matrices, $v$ the velocity of the Dirac part of the fermion dispersion which is linear in momentum.  In the hexagonal warping term   $k_\pm=k_x\pm i k_y$ with $k_x$, $k_y$ momentum components in the surface plane.  The $\Delta$ is a gap and $E(k)=\frac{\hbar^2 k^2}{2m}\equiv E_0 k^2$ a quadratic dispersion.  We are interested in the case when the first Dirac term in Eq.~\ref{eqn:H} is dominant and $E_0$ is by comparison small.  To be specific we will take $v=2.8 \times 10^5$m/s and $m$ equal to the electron mass $m_e$, which we refer to as $E_0=1$ in the appropriate units of $\frac{\hbar^2}{2m_e}$.  These values are illustrative only.  For the specific case of Bi$_2$Te$_3$ for example, $v=4.3 \times 10^5$m/s and $m=0.9m_e$.\cite{li:2013} A fit to angular resolved photoemission data on Bi$_2$Te$_3$ by Fu gave a value of $\lambda\approx 250$~meV\AA$^3$ which sets the order of magnitude for this coupling.\cite{fu:2009,li:2013:2}  The energies are given by
\begin{equation}\label{eqn:ek}
E_s(\vect{k})=E(k) + s\sqrt{\hbar^2 v^2 k^2 +\left(\Delta+\lambda \{ k_x^3 -3k_x k_y^2 \}\right)^2 }.
\end{equation}
 The directionally dependent part in Eq.~(\ref{eqn:ek}) can be rewritten in terms of the polar angle $\theta_{\vect{k}}$ for the vector $\vect{k}$ as
\begin{equation}
\Delta(k,\theta_{\vect{k}})\equiv \Delta +\lambda k^3 \cos(3\theta_{\vect{k}}).
\end{equation}
The eigenvectors are then dependent on the magnitude and direction of the momentum $\vect{k}$, as well as the band index, $s=\pm1$ and are given by
\begin{widetext}
\begin{equation}\label{eqn:u}
u(\vect{k},s) = \hbar v k \frac{\begin{pmatrix}
1 , \frac{1}{\hbar v k^2}\left[ \Delta(k,\theta_{\vect{k}}) -s \sqrt{\hbar ^2 v^2 k^2 + \Delta(k,\theta_{\vect{k}})^2 }\right] (-ik_x +k_y)
\end{pmatrix}^\mathsf{T}}{\sqrt{\hbar ^2 v^2 k^2 + (\Delta(k,\theta_{\vect{k}}) - s\sqrt{\hbar ^2 v^2 k^2 + \Delta(k,\theta_{\vect{k}})^2 }   )^2 }}.
\end{equation}
\end{widetext}

\begin{figure}
  \begin{center}
  \includegraphics[width=0.95\linewidth]{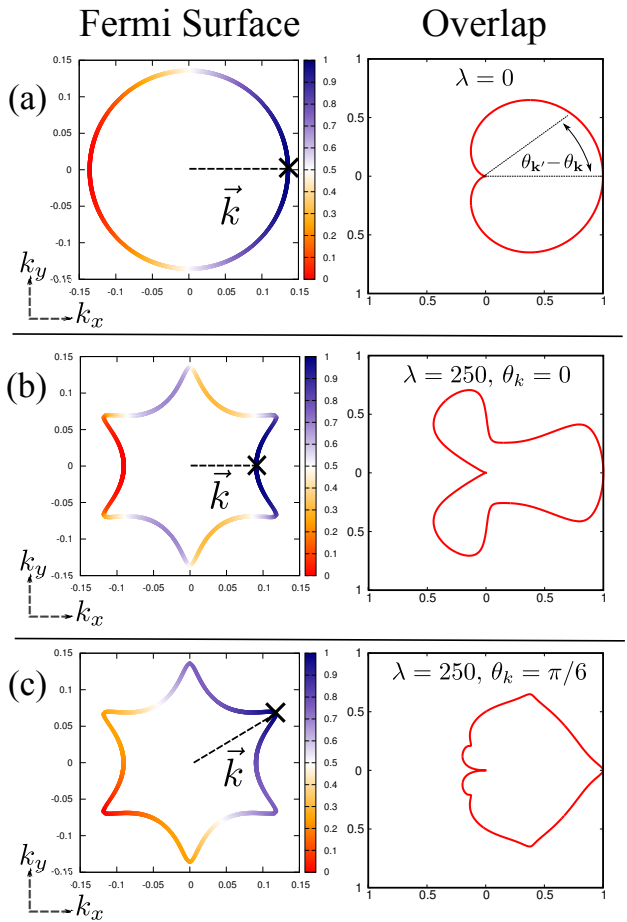}
  \end{center}
  \caption{\label{fig:polar}(Color online) 
  Overlap between states at the Fermi level for $\mu=0.25$~eV with initial momentum $\vect{k}=(k_F,\theta_{\vect{k}})$ and final $\vect{k^\prime}=(k_F, \theta_{\vect{k^\prime}})$ measured in \AA$^{-1}$.  Left: the momenta of $\vect{k^\prime}$ as a function of $k_x$ and $k_y$ with a color scale which represents the overlap $\left| \langle u_{\vect{k^\prime}} | u_{\vect{k}} \rangle  \right|^2$ for an initial momentum $\vect{k}$ marked in each frame by an `$\boldsymbol{\times}$'.  Right: shows a corresponding polar plot of $\left| \langle u_{\vect{k^\prime}} | u_{\vect{k}} \rangle  \right|^2$ as a function of $\theta_{\vect{k^\prime}}-\theta_{\vect{k}}$.  (a) $\lambda=0$, Dirac case. (b) $\lambda=250$~meV\AA$^3$, for initial momentum along a hexagonal side, $\vect{k}=(k_F,\theta_{\vect{k}}=0)$.  (c) $\lambda=250$~meV\AA$^3$, for initial momentum along a hexagonal vertex, $\vect{k}=(k_F,\theta_{\vect{k}}=\pi/6)$.}
\end{figure}

To orient the reader, we first discuss the impact of these various interaction terms on the electronic dispersion, which we sketch schematically in Fig.~\ref{fig:hex}.  In order from left to right we show the dispersions for pure Dirac, a hexagonal warping term, followed by an additional gap and the final frame illustrates the effect of a sub-dominant Schr\"odinger term without a gap. 
The polarization, $\Pi(\vect{q},\omega)$, is the fundamental quantity needed to describe the dielectric properties of an electronic system.  For our model Hamiltonian it can be written as
\begin{widetext}
\begin{equation}\label{eqn:pol}
\Pi(\vect{q},\omega)=\sum\limits_{ss^\prime}\int\frac{d\vect{k}}{(2\pi)^2} [ f(E_s(\vect{k})) -f(E_{s^\prime}(\vect{k}+\vect{q}))] |\langle u(\vect{k},s) |  u(\vect{k}+\vect{q},s^\prime) \rangle |^2 \left[ \frac{1}{\omega-E_s(\vect{k})+E_{s^\prime}(\vect{k}+\vect{q}) +i\Gamma}  \right]
\end{equation}
\end{widetext}
where $f$ is the Fermi function, $\Gamma$ is a small intrinsic scattering rate, and the overlap matrix element squared represents scattering from the state $| \vect{k} , s \rangle$ to the state $\langle \vect{k}+\vect{q} , s^\prime |$, and will depend not only on the magnitude of $\vect{q}$ but also on its direction with respect to the two dimensional surface state Brillouin zone as a result of the hexagonal warping which introduces a new element of complexity not encountered in the pure Dirac spectrum.

The overlap between states at the Fermi level is known to be essential to the low energy physics in graphene.  An example of this is the well known chirality induced removal of backscattering processes which is understood by the overlap shown in Fig.~\ref{fig:polar}(a).  Here there is a strong preference for forward scattering processes for states scattering within the same cone which we illustrate in the left hand column using blue(red) color scale for strong(weak) scattering amplitudes which are given explicitly in the right hand column for an electron with initial momentum along $k_x$ or $\theta_{\vect{k}}=0$.  The inclusion of hexagonal warping complicates this in a non-trivial manner.  There is a very different angular overlap depending on the direction of the initial momentum, $\vect{k}$, of the electron.  If the initial momentum is along the side of the hexagon, as shown in Fig.~\ref{fig:polar}(b) for $\theta_{\vect{k}}=0$, then for relevant values of $\lambda$ there is now a strong preference to scattering through an angle of $2\pi/3$ which results in enhanced scattering between alternating sides of the hexagonal Fermi surface.  If instead the initial momentum is along the corner of the hexagon, shown in Fig.~\ref{fig:polar}(c) for $\theta_{\vect{k}}=\pi/6$, then the scattering amplitude is not significantly different from the case of a circular Fermi surface.

\section{Numerical Results}\label{sec:num}
The inclusion of the anisotropic hexagonal term poses a significant hurdle to the analytic evaluation of Eq.~(\ref{eqn:pol}).   Here we instead describe the full Hamiltonian under a single framework which can be evaluated numerically.  Because of the large number of parameters, what we present is by no means an exhaustive description of possible results, but instead a minimal set to provide a qualitative description of the effects of each term in the Hamiltonian.  In addition, there exist some analytic results which have been previously evaluated for a pure Dirac system,\cite{wunsch:2006, barlas:2007} as well as a system with a gap term.\cite{scholz:2012}
In order to qualitatively compare with these results, we proceed numerically in the clean limit of scattering $\Gamma \to 0^+ <<\hbar vq$.  We will see later that this is an essential element to correctly describing the plasmon dispersions from numerics. 
In the case of graphene it is common to scale momenta by the value at the Fermi level, and frequency by the chemical potential.  Here we do not do this for several reasons.  One reason is that the inclusion of anisotropy due to hexagonal warping creates an angularly dependent value of the Fermi momentum, $k_F(\theta)$.  Further, changing each parameter changes the relative Fermi level. We therefore maintain a single set of unscaled units, so that these differences can be seen in our axes labels.  With this in mind we proceed with an experimentally relevant chemical potential of $\mu=250$~meV.

\begin{figure}
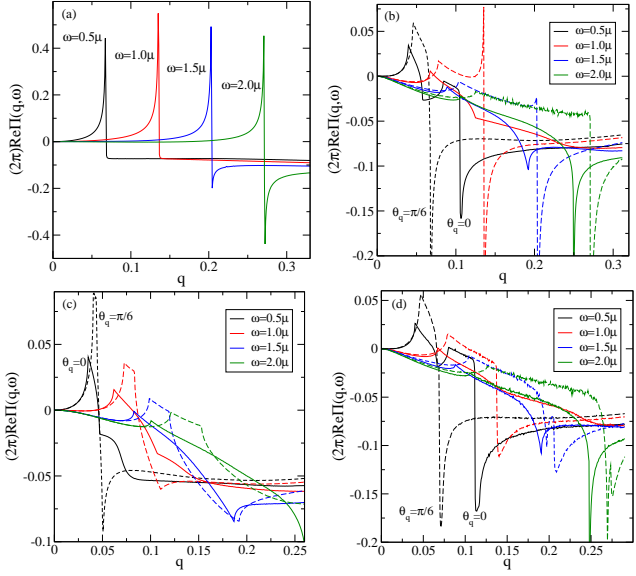

\centering
\begin{subfigure}{.49\linewidth}
  \centering
  \includegraphics[width=.95\linewidth]{realpol-dirac.eps}
   \label{fig:repol:a}
\end{subfigure}%
\begin{subfigure}{.49\linewidth}
  \centering
  \includegraphics[width=.95\linewidth]{realpol-l50.eps}
    \label{fig:repol:b}
\end{subfigure}\\
\begin{subfigure}{.49\linewidth}
  \centering
  \includegraphics[width=.95\linewidth]{realpol-l50-E01.eps}
    \label{fig:repol:c}
\end{subfigure}
\begin{subfigure}{.49\linewidth}
  \centering
  \includegraphics[width=.95\linewidth]{realpol-l50-Dp05.eps}
   \label{fig:repol:d}
\end{subfigure}
\caption{(Color Online) Real part of polarization function ${\rm Re}\Pi(q,\omega)$ for four values of $\omega$ as a function of $q$ in units of \AA$^{-1}$.  (a) graphene case for comparison. (b) including hexagonal warping of $\lambda =50$~meV\AA$^3$, $E_0 =0$ and $\Delta =0$ (c)$\lambda =50$~meV\AA$^3$, $E_0 =1.0$ and $\Delta=0$. (d)$\lambda =50$~meV\AA$^3$, $E_0 =0$ and $\Delta=$50~meV. $\theta_q=0$ and $\theta_q=\pi/6$ are shown as solid and dashed lines respectively in each frame. }
\label{fig:repol}
\end{figure}

\begin{figure}
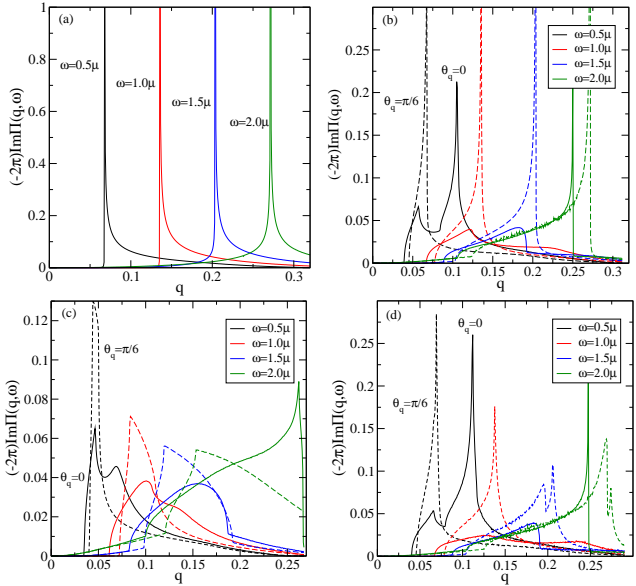

\centering
\begin{subfigure}{.49\linewidth}
  \centering
  \includegraphics[width=.95\linewidth]{imagpol-dirac.eps}
    \label{fig:impol:a}
\end{subfigure}%
\begin{subfigure}{.49\linewidth}
  \centering
  \includegraphics[width=.95\linewidth]{imagpol-l50.eps}
  \label{fig:impol:b}
\end{subfigure}\\
\begin{subfigure}{.49\linewidth}
  \centering
  \includegraphics[width=.95\linewidth]{imagpol-l50-E01.eps}
  \label{fig:impol:c}
\end{subfigure}
\begin{subfigure}{.49\linewidth}
  \centering
  \includegraphics[width=.95\linewidth]{imagpol-l50-Dp05.eps}
   \label{fig:impol:d}
\end{subfigure}
\caption{(Color Online)  Imaginary part of polarization function ${\rm Im}\Pi(q,\omega)$ for four values of $\omega$ as a function of $q$ in units of \AA$^{-1}$.  (a) graphene case for comparison. (b) including hexagonal warping of $\lambda =50$~meV\AA$^3$, $E_0=0$ and $\Delta=0$ (c)$\lambda =50$~meV\AA$^3$, $E_0=1.0$ and $\Delta=0$. (d)$\lambda =50$~meV\AA$^3$, $E_0=0$ and $\Delta=$50~meV. $\theta_q=0$ and $\theta_q=\pi/6$ are shown as solid and dashed lines respectively in each frame. }
\label{fig:impol}
\end{figure}

In Fig.~\ref{fig:repol} we evaluate numerically the real part of the polarization function of Eq.~(\ref{eqn:pol}) for several choices of parameters $\lambda$, $\Delta$, and $E_0$.    Fig.~\ref{fig:repol}(a) is for comparison with previous analytic work on graphene and shows the real part of $\Pi(\vect{q},\omega)$ in units of \AA$^{-2}$/eV as a function of $q$.
  These numerical results (scaling aside) agree precisely with analytic results presented in Fig.3(c) of Kotov et al\cite{kotov:2012} and elsewhere.\cite{leblanc:2011, wunsch:2006}
Results are presented in the same format for four values of $\omega/\mu=0.5$ (black), 1.0 (red), 1.5 (blue) and 2.0 (green).  Frame (b) is for the same values of $\omega$ but now includes hexagonal warping with $\lambda=50$~meV\AA$^3$.  An important new feature is that now the polarization depends not only on the magnitude of $\vect{q}$ but also depends on direction.  Results for $\theta_{\vect{q}}=0$ are represented with solid lines while for $\theta_{\vect{q}}=\pi/6$ we use dashed lines with similar color coding.  From this it is clear that there can be a great deal of anisotropy which is directly related to the warping term.  Differences are small however in the limit of $\vect{q}$ going to zero, since in this region the solid and dashed curves overlap. As $\vect{q}$ is increased the departures between the two can be very significant; not just small quantitative changes, but large ones which change the qualitative behavior.  
As an example the dashed black curve for $\theta_{\vect{q}}=\pi/6$ at $\omega/\mu=0.5$ shows a single maximum around $q\cong 0.05$ while for $\theta_{\vect{q}}=0$ the peak is split into two and the second peak in the doubled maxima structure is shifted to the right to higher values of $q$.
Comparison with the curves in frame (a) for the pure Dirac case shows that new structures are introduced by the hexagonal warping term.  In particular the most prominent maximum in the graphene case is now less prominent and essentially spread over a range of momenta because of the anisotropy that is introduced.
The splitting of the two angles at low $\omega$ and small $q$ can be understood in our description of the overlaps in Fig.~\ref{fig:polar}.  In the $\theta_{\vect{q}}=0$ case the curves rise until a momentum which begins to sample the adjoining hexagonal sides (light red in Fig.~\ref{fig:polar}(b) where the overlap is small, and then has a second rise once $q$ is large enough to access the opposing hexagonal faces (light blue in Fig.~\ref{fig:polar}(b)).  This behaviour is not seen for $\theta_{\vect{q}}=\pi/6$, which has overlaps which are more Dirac-like.  We will see this behaviour, that the $\theta_{\vect{q}}=\pi/6$ direction is more Dirac like, for various quantities throughout this paper.
  Turning next to Fig.~\ref{fig:repol}(c) we show that the addition of a small Schr\"odinger piece to the Hamiltonian ($E_0=1$) leads to further quantitative changes.  In particular, for $\theta_{\vect{q}}=0$ the solid black curve has lost its second peak which has been replaced by a weak shoulder  around $q\approx 0.05$\AA$^{-1}$ instead.  Anisotropy with angle $\theta_{\vect{q}}$ remains most significant in the intermediate $q$ range.  
Frame (d) is for $\lambda=50$~meV\AA$^3$ with $E_0=0$ but now we have also included a gap of $\Delta= 50$~meV.  Comparing with (b) we note that the introduction of a gap leads to many changes particularly in the height of the peaks.  We note that in all cases, there is no indication of angular anisotropy at small $q$ and that the most notable qualitative changes are seen in the dashed (blue and green) curves at the higher values of frequency  and momentum transfer.  Here the single minima at large $q$ in (b) splits into two in (d) due to the presence of the gap for $\theta_{\vect{q}}=\pi/6$.  This does not occur along $\theta_{\vect{q}}=0$.
Corresponding results for the imaginary part of the polarization function ${\rm Im}\Pi(\vect{q},\omega)$ vs $q$ for the same four values of $\omega$ are presented in Fig.~\ref{fig:impol}.  Comparing our new results shown in Fig.~\ref{fig:impol}(b) which include warping contributions against the pure Dirac case of Fig.~\ref{fig:impol}(a) we emphasize two features.  First the onset of non-zero damping is shifted to lower energies and momenta by the anisotropic warping term and now also depends on the direction of $\vect{q}$.  Secondly as we have already noted in discussion of Fig.~\ref{fig:repol} the sharp peak of the Dirac case has become spread over a range of values of $q$ and as can be seen in the black curve of frame (b) the structure is split into two pieces.  
Quantitative modifications arise when a quadratic piece is included.  For example, in Fig.~\ref{fig:impol}(c) the first peak in the black solid curve is higher than the second in contrast to the $E_0=0$ case where this is reversed.
\begin{figure}
\centering
 \includegraphics[width=0.85\linewidth]{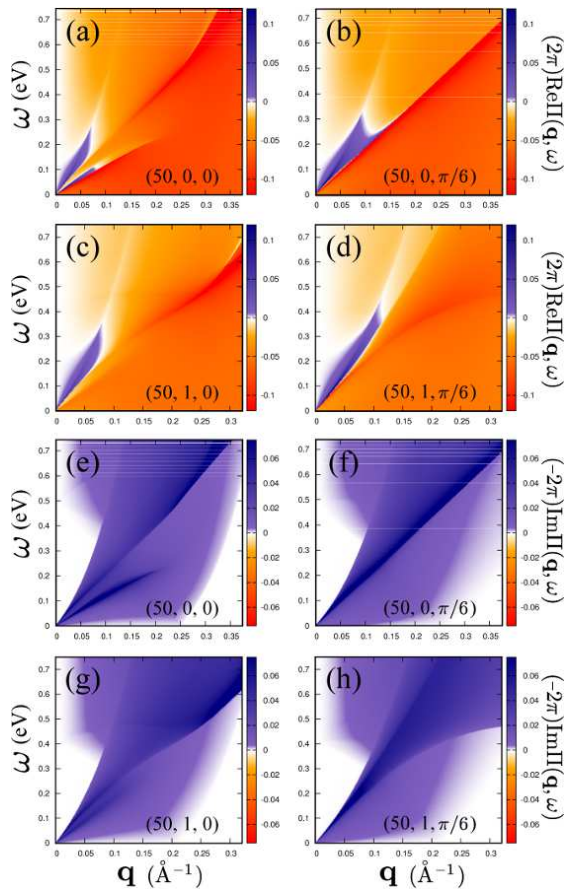}
    \caption{(Color online) Color plots of the real and imaginary parts of the polarization function for parameters labelled in each frame in the format $(\lambda, E_0, \theta_{\vect{q}})$. }
  \label{fig:6}
\end{figure}%

In order to properly understand the source of these complicated new features, we plot in Fig.~\ref{fig:6} a set of full color-map plots of our computed polarization function on a fine grid of $q$ and $\omega$. The first four top frames give our results for the real part of the polarization function.  Of these, the upper two frames include no Schr\"odinger piece ($E_0=0$) but show two directions of the scattering $\vect{q}$; (a) for $\theta_{\vect{q}}=0$, (b) for $\theta_{\vect{q}}=\pi/6$.  
In the absence of anisotropy these would be identical.  However, we note a great deal of anisotropy for finite but small values of $|\vect{q}|$.  Also the boundary between negative and positive regions of ${\rm Re}\Pi(\vect{q},\omega)$ in (a) merge into a single region in (b). The next row, frames (c) and (d), includes a small Schr\"odinger term of $E_0=1$ in addition to the dominant Dirac term and the hexagonal warping term.  This introduces further changes, particularly in the blue region at small $q$ and $\omega$ which corresponds to the plasmon region.
In all cases this plasmon region, where ${\rm Re}\Pi(\vect{q},\omega) >0$, is extremely restricted and does not extend far along the $\omega=\hbar v q$ line as it would in the pure Dirac case.  We will see later that this has significant consequences on the plasmon dispersions.
The most striking change due to the inclusion of a Schr\"odinger term is the complete removal of the splitting feature which is seen along $\theta_{\vect{q}}=0$ in Fig.~\ref{fig:6}(a).
The lower four frames give the corresponding imaginary part of the polarization and also show quantitative changes with value of $E_0$ and angle $\theta_q$.  In particular, the boundaries of the particle-hole continuum change with 
$\lambda$, $E_0$, and $\theta_{\vect{q}}$ .
These differences are seen in (e) and (f) for the imaginary part. 
Also, it is in the imaginary part of the polarization that the regions in $q$ and $\omega$ space are most evident.  Examining first Fig.~\ref{fig:6}(e), we see at low $\omega$ first the Pauli blocked region, then the intraband region as $q$ is increased, as well as the edge of the intraband piece which occurs beyond $q=2k_F$ for $\omega=0$.  In this case, unlike the pure Dirac case, the intraband region has a non-linear edge due to the hexagonal warping.  One can also see a split residual linear feature as was pointed to in Fig.~\ref{fig:6}(a) and accentuated there as the lower blue region where ${\rm Re}\Pi(\vect{q},\omega)>0$.
Fig.~\ref{fig:6}(f) shows the $\theta_{\vect{q}}=\pi/6$ direction where the imaginary part is again much more Dirac-like.  The primary difference between directions of $\vect{q}$ is the lack of a split feature in the intraband continuum.  It is clear that with $\lambda \neq 0$ the boundaries of the inter and intraband particle-hole excitations shift with angle $\theta_{\vect{q}}$.  These changes will have an effect on the stability of the plasmons because the plasmons remain undamped only where ${\rm Im}\Pi(\vect{q},\omega)=0$, shown in white.
    We will see in Sec.~\ref{sec:anal} how these considerations restrict the extent of the plasmon dispersion with variation in hexagonal warping, and other terms in the Hamiltonian.

\begin{figure}
\centering
 \includegraphics[width=0.85\linewidth]{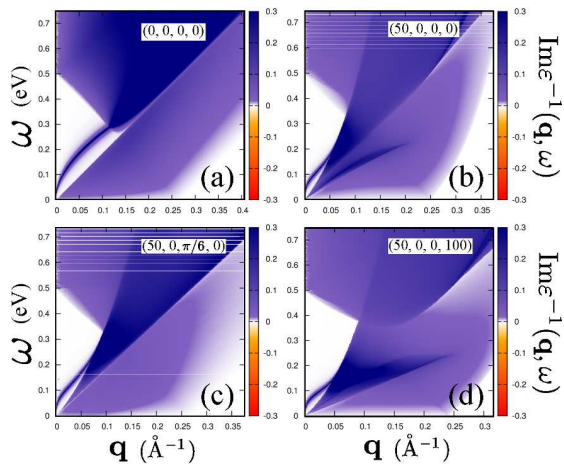}
    \caption{(Color online) Color plots of the imaginary part of the inverse dielectric function ${\rm Im}\upvarepsilon^{-1}(q,\omega)$ for parameters labelled in each frame in the format $(\lambda, E_0, \theta_{\vect{q}}, \Delta)$ and $\hbar v \alpha_{FS} =5$. }
  \label{fig:7}
\end{figure}

We next turn to the effects of $\lambda$ and $E_0$ on the loss function, ${\rm Im}\upvarepsilon^{-1}(\vect{q},\omega)$, which is a measurable quantity. The full dielectric function of the surface states is given by
\begin{equation}
\upvarepsilon(\vect{q},\omega)=1 -V(\vect{q})\Pi(\vect{q},\omega),
\end{equation}
where  $V(\vect{q})=\frac{2\pi e^2}{\epsilon_0 |\vect{q}|}$ is the Coulomb potential and $\epsilon_0$ is the effective dielectric constant of the medium. 
Results for ${\rm Im}\upvarepsilon^{-1}(\vect{q},\omega)$ are presented in Fig.~\ref{fig:7}.  Fig.~\ref{fig:7}(a) provides results for the pure Dirac spectrum as a comparison case.   One can see clearly the boundaries of the intra- and interband parts of the particle-hole continuum which corresponds to the shaded regions.  The regions which are white are for values of $q$ and $\omega$ where there are no particle-hole excitations due to Pauli blocking at $T=0$.
On the lower left corner of the figure we see a prominent blue curve which is the plasmon dispersion curve.  Here this curve has a width because the use of a small residual scattering rate results in a small but finite value of ${\rm Im}\Pi(\vect{q},\omega)$ which is then susceptible to the plasmon pole, creating a very sharp peak in $\upvarepsilon^{-1}(\vect{q},\omega)$.  At higher values of $\omega$ and $q$ the plasmon branch enters the particle-hole continuum and becomes Landau damped.  The inclusion of warping shifts the boundaries of the particle-hole continuum and in the case where $\theta_{\vect{q}}=0$ in Fig.~\ref{fig:7}(b) the plasmons become damped at lower $q$ and $\omega$ relative to the $\lambda=0$ case.  Further changes in the plasmon dispersion are seen in the lowest left frame which is for $\theta_{\vect{q}}=\pi/6$.  Although this case includes the same warping contribution as in (b) it resembles instead the results of (a) for the pure Dirac case except for the non-linear intraband onset.  Finally, Fig.~\ref{fig:7}(d) includes a gap of 100~meV.
The dielectric function of a gapped Dirac spectrum has been previously examined.\cite{scholz:2012}  Here, the inclusion of hexagonal warping causes the intraband piece to be non-linear resulting in no actual gap between the intraband and interband regions of the particle-hole continuum.  This severely disrupts the long lived plasmons regardless of the direction of $\vect{q}$.

\section{Long Wavelength Limit of Plasmon Dispersion}\label{sec:anal}
Prior to considering numerics it is useful to derive in the limit of $q\to 0$ an analytic expression for the plasmon dispersion,\cite{dassarma:2009, wunsch:2006, kechedzhi:2013} $\omega_p(\vect{q})$, which follows from the solution to
\begin{equation}\label{eqn:plasmon}
1=V(\vect{q}){\rm Re}\Pi(\vect{q},\omega_p).
\end{equation} 
We can use such analytic work to guide our expectations for how various terms modify the plasmon dispersion. For definiteness we take the chemical potential to fall in the upper Dirac-cone.  For $q \to 0$ the low energy plasmon dispersion will then depend only on the intraband piece for the case where $s=s^\prime=1$.  In this case
\begin{widetext}
\begin{equation}\label{eqn:analpol}
{\rm Re} \Pi(\vect{q},\omega)= \int \limits_0^{\Lambda} \frac{k dk}{2\pi}\int \limits_0^{2\pi} \frac{d\theta}{2\pi} \frac{f(E_+(\vect{k^\prime})) -f( E_+(\vect{k}) )  }{ [E_+(\vect{k^\prime}) - E_+(\vect{k})]^2 -\omega^2 }  [E_+(\vect{k^\prime}) - E_+(\vect{k})],
\end{equation}
\end{widetext}
where $\Lambda$ is a cutoff, $\vect{k^\prime}=\vect{k}+\vect{q}$ and where we have used the symmetry ${\rm Re}\Pi(\vect{q},\omega)={\rm Re}\Pi(\vect{q},-\omega)$.  To lowest order in $q$ we write $E_+(\vect{k}+\vect{q})\cong q\beta(k,\theta, \alpha)$ where $\theta$ defines the direction of $\vect{k}$, $\alpha$ defines the angle of $\vect{q}$ and $\beta(k,\theta, \alpha)$ is given in Eq.~(\ref{eqn:beta}).  When the polarization does not depend on $\alpha$ by symmetry $\vect{q}$ can be taken along the $k_x$ axis and only $\theta$ remains which is integrated over in Eq.~(\ref{eqn:analpol}).   The Fermi factors in the numerator of this equation give a factor of $q \beta(k, \theta, \alpha) \frac{\partial f(E_+(\vect{k}))}{\partial E_+(\vect{k})}$ and $E_+(\vect{k}+\vect{q})- E_+(\vect{k})$ gives another factor of $q \beta(k,\theta,\alpha)$.  This last factor also appears in the denominator but it can be dropped relative to $\omega$ as it is of order $q^2$ and $\omega$ will turn out to be of order $q$.  Putting all of this together we obtain
\begin{equation}
{\rm Re}\Pi(\vect{q},\omega) =\frac{1}{2\pi}\int \limits_0^{\Lambda} k dk\int \limits_0^{2\pi} \frac{d\theta}{2\pi}  \left( -\frac{\partial f(E_+(\vect{k}))}{\partial E_+(\vect{k})} \right) \frac{q^2 \beta^2(\vect{k},\theta,\alpha)}{\omega^2}
\end{equation}
which goes like $q^2$ and holds at any temperature $T$.  In the limit of zero temperature the derivative of the Fermi function becomes a Dirac delta function $\delta(E_+(\vect{k}) -\mu)$.  For values of $\mu$ much less than the band width which is the case of interest here the cut-off, $\Lambda$, on $|\vect{k}|$ is of no consequence and the delta function for a given angle $\theta$ will contribute only for $k=k_c(\theta)$ at energy  $E_+(\vect{k})=E_+(k_c(\theta),\theta)=\mu$.  Doing the $k$ integration first gives
\begin{equation}\label{eqn:9}
{\rm Re}\Pi(\vect{q},\omega) = \frac{q^2}{\omega^2} \int \limits_0^{2\pi} \frac{d\theta}{(2\pi)^2} \frac{k_c(\theta) \beta^2(k_c(\theta),\theta,\alpha))}{  \left(\frac{d E_+(k, \theta)}{dk}\right)_{k=k_{c}(\theta)  }},
\end{equation}
where $\beta$ is a function defined in the appendix Eq.~(\ref{eqn:beta}).

\begin{figure}
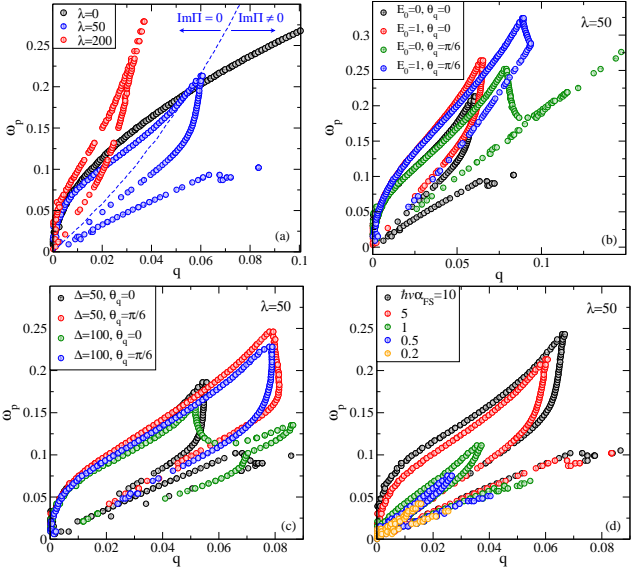

\centering
\begin{subfigure}{.49\linewidth}
  \centering
  \includegraphics[width=.95\linewidth]{a2-new.eps}
\end{subfigure}%
\begin{subfigure}{.49\linewidth}
  \centering
  \includegraphics[width=.95\linewidth]{b2.eps}
\end{subfigure}\\
\begin{subfigure}{.49\linewidth}
  \centering
  \includegraphics[width=.95\linewidth]{c2.eps}
\end{subfigure}
\begin{subfigure}{.49\linewidth}
  \centering
  \includegraphics[width=.95\linewidth]{d2.eps}
\end{subfigure}
\caption{(Color Online) Numerical extraction of $\omega_p(q)$ from Eqs.~(\ref{eqn:pol}) and (\ref{eqn:plasmon}) for: (a) variation in hexagonal warping strength, $\lambda=50$, and 200~meV\AA$^3$ along $\theta_{\vect{q}}=0$ (b) $\lambda=50$~meV\AA$^3$ and inclusion of $E_0$ term for $\theta_{\vect{q}}=0$ and $\pi/6$, (c) $\lambda=50$~meV\AA$^3$  and a gap of $\Delta=50$, 100~meV, (d) $\lambda=50$~meV\AA$^3$ for variation in coupling strength $\alpha_{FS}$ with $\theta_{\vect{q}}=0$ and $E_0=0$.  In (a-c) a value of $\hbar v \alpha_{FS} =5$ is taken for illustrative purposes.}
\label{fig:wp}
\end{figure}

\begin{figure}
\centering
 \includegraphics[width=0.95\linewidth]{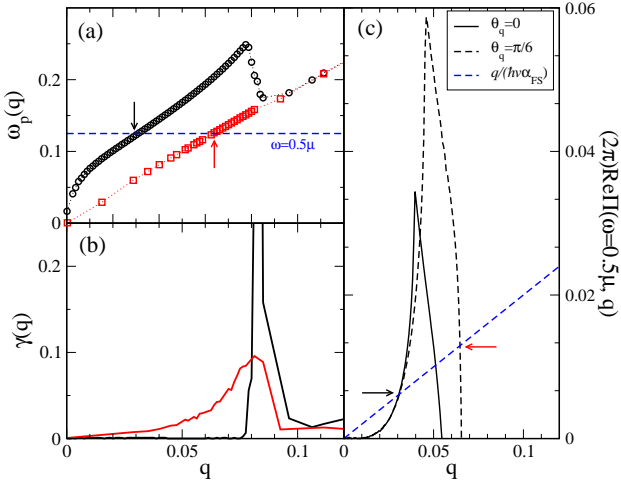}
    \caption{(Color online) Examination of multiple plasmon poles for the single case of $\lambda=50$~meV\AA$^3$, $E_0=0$, $\Delta=0$, and for $\hbar v \alpha_{FS} =5$.   (a)  Separate plasmon branches, $\omega_p(q)$, shown in red/black (circles/squares) for $\theta_{q}=\pi/6$. (b) The corresponding  plasmon decay rate, $\gamma(q)$, for each plasmon branch in frame (a). (c) A cut of the real part of the polarization, ${\rm Re}\Pi(\omega=0.5\mu,q)$, given by the dashed blue line in (a).  Black and Red colored arrows in frames (a) and (c) mark the corresponding plasmon poles, and their multiple solutions. }
  \label{fig:new}
\end{figure}

As we advertised, this will result in  $\omega_p \propto \sqrt{q}$ which justifies the approximations made.  An integral over the angles of $\vect{k}$ remains to be performed.  For pure Dirac dispersion $E_+(\vect{k})=\hbar vk$, and the  $\beta(k_c(\theta),\theta, \alpha)$ function reduces to $\hbar v \cos\theta$, $k_c=\mu/(\hbar v)$ and $\left(\frac{dE_+(k,\theta)}{dk}\right)_{k=k_c}=\hbar v$ such that
\begin{equation}\label{eqn:wp}
\omega_p(q)=\sqrt{\frac{e^2\mu}{2\epsilon_0}}\sqrt{q}=\sqrt{\frac{\alpha_{FS} \mu \hbar v q}{2}}
\end{equation}
which agrees with the known result for graphene where $\alpha_{FS}=e^2/(\epsilon_0 \hbar v)$ is the effective screened fine structure constant.\cite{wunsch:2006,jang:2008}  In that case the factor of 2 in Eq.~(\ref{eqn:wp}) is in the numerator because of the spin/valley degeneracy which is present in  graphene but is not present in a topological insulator.

It is interesting to examine other simple limits complimentary to our numerical work.  For example we can take a Dirac term plus a gap, in which case $\beta(k,\theta,\alpha)=k\cos\theta \frac{\hbar^2v^2}{\sqrt{\hbar^2v^2k_c^2 + \Delta^2}}$ and $\sqrt{\hbar^2v^2k_c^2 + \Delta^2}=\mu$ which leads to
\begin{equation}\label{eqn:wpd}
\omega_p(q)=\sqrt{\frac{e^2\mu}{2\epsilon_0} q\left[1-\left(\frac{\Delta}{\mu}\right)^2\right] }.
\end{equation}
The magnitude of the gap is assumed to be much less than $\mu$, and has the effect of reducing the slope of the $\sqrt{q}$ dependence of $\omega_p(q)$.  Next we take a Dirac term plus a Schr\"odinger term.  Assuming $E_0$ is small gives the lowest order correction
\begin{equation}\label{eqn:wpE}
\omega_p(q)=\sqrt{\frac{e^2 \mu}{2\epsilon_0} q \left[1 + \frac{2E_0 \mu}{(\hbar v)^2}\right] }.
\end{equation}
In this case the slope of the $\sqrt{q}$ dependence of the plasmon dispersion is increased.  The case of Dirac plus hexagonal warping involves more complex, but still straightforward algebra.  Some of the necessary work is described in the appendix.  The result is:
\begin{equation}\label{eqn:wpl}
\omega_P(q)= \sqrt{ \frac{e^2 \mu}{\epsilon_0} q \left[ \frac{1}{2} + \left(\frac{\lambda}{\hbar v}\right)^2\left(\frac{\mu}{\hbar v}\right)^4 h(\alpha) \right]   }
\end{equation}
where $h(\alpha)$ is a function of $\alpha$ integrated over $\theta$(see Appendix) which surprisingly comes out to be a constant value of $h(\alpha)=1/2$.

To obtain Eq.~(\ref{eqn:wpl}) we have assumed that $\lambda$ was small and worked to lowest order.  In general, $\alpha$ will not drop out of $h(\alpha)$ and the plasmon dispersion will depend on the angle of $\vect{q}$ relative to the Brillouin zone axis.  There is no simple analytic formula that covers the higher order case however, and we need to proceed numerically as we will do next.

Numerical results for the plasmon dispersion are presented in Fig.~\ref{fig:wp}.  In Fig.~\ref{fig:wp}(a) we explore the variation in $\omega_p(\vect{q})$ vs $|\vect{q}|$ for the case $\theta_{\vect{q}}=0$, $E_0=0$ when the hexagonal warping term is increased.  The black circles are for $\lambda=0$, the pure Dirac case, and are for comparison.  The blue circles apply for $\lambda=50$~meV\AA$^3$ and the red for $\lambda=200$~meV\AA$^3$.  It is clear that these data conform with our expectation, based on Eqn.~(\ref{eqn:wpl}), that the slope of the plasmon dispersion curve increases with $\lambda$ in the long wavelength limit.  Differences between black, blue and red data rapidly increase with increasing scattering momentum, $|\vect{q}|$.  Several other features are to be noted.  The red and blue curves first increase out of $q=0$ come to a maximum and then fold back to smaller values of $\omega$.  
For a single case of $\lambda=50$ we have marked the intraband transition line (dashed-blue line) where ${\rm Im}\Pi(q,\omega)$ goes from zero to a finite value.  We can see that the uppermost line is undamped until it passes into the particle-hole continuum, where it will become damped.  This curve then folds back as the dispersion gets close to a sign change in the real part of the polarization function, shown for example in the color-plots of Fig.~\ref{fig:6}(a).
  For the blue curve there is even a third region which is entirely within the Landau damped region related to the onset of a residual linear intraband transition.  While these data points correspond to a solution of Eq.~(\ref{eqn:plasmon}), they do not represent undamped plasmons as they overlap with the particle-hole continuum. In Fig.~\ref{fig:7}(a) we saw a single plasmon dispersion in the Pauli-blocked region which continues into and merges with the interband particle-hole continuum.  This contour in Fig.~\ref{fig:7}(a) corresponds to the black circles of Fig.~\ref{fig:wp}(a).  In Fig.~\ref{fig:7}(b) the hexagonal warping increased the region of the particle hole continuum so that the plasmon line now ends at smaller values of $\omega$ and $q$.  This is shown in Fig.~\ref{fig:wp}(a) for variation in hexagonal warping strength.  We see in all cases the $\sqrt{q}$ behavior in the uppermost branch of $\omega_p$ which comes from intraband transitions as in our analytic work.

In Fig.~\ref{fig:wp}(b) we compare results for $\lambda=50$~meV\AA$^3$ with and without a Schr\"odinger contribution and for the two key angles of $\theta_{\vect{q}}=0$ and $\pi/6$.
The slopes out of $q=0$ are only weakly modified by angle as expected from Eq.~(\ref{eqn:wpl}), but show strong dependence on the inclusion of a Sch\"odinger piece ($E_0=1$) in agreement with our approximate but analytic result in Eq.~(\ref{eqn:wpE}).  The numerical agreement between angular directions confirms our assertion from Eq.~(\ref{eqn:wpl}) that the slope of the plasmon dispersion is unaffected by the anisotropy of the hexagonal warping for small but relevant values of $\lambda$, and that this remains true even for larger values of $\lambda$.  
Fig.~\ref{fig:wp}(c) gives our results for two values of the gap, $\Delta=50$ and 100~meV again for the two relevant angles in the presence of weak hexagonal warping of $\lambda=50$~meV\AA$^3$.  We can see that the value of the gap has very little effect on the uppermost plasmon solutions, but a more noticeable impact on the lower branches which occur at the intraband/interband continuum boundaries.  For large gaps these boundaries can be substantially modified.  As predicted in our analytics of Eq.~(\ref{eqn:wpd}) we see that the increase in $\Delta$ results in a small reduction of the slope out of $q=0$.
The final frame, Fig.~\ref{fig:wp}(d), shows how the plasmon dispersions are changed when the dielectric constant of the environment is changed, here written in terms of the effective fine structure and Fermi velocity factors, $ \hbar v \alpha_{FS}$.  Decreasing this quantity decreases the slope of the plasmon out of $q=0$ in agreement with all analytic results.  As the coupling strength, $\alpha_{FS}$ is reduced, we see that the upper and lower plasmon branches merge and shrink towards lower $q$ and $\omega$.

Finally we examine the existence of separate plasmon branches in Fig.~\ref{fig:new}.  Here we pull a single case from Fig.~\ref{fig:wp}(b), at the angle of $\theta_{\vect{q}}=\pi/6$ (green points), which contains two branches, one with a square root dependence, and a second linear plasmon branch at small $q$.  We identify these branches in Fig.~\ref{fig:new}(a) as black and red (circles and squares) respectively.  To illustrate the truly distinct nature of these branches, we plot in Fig.~\ref{fig:new}(b) the corresponding plasmon decay rate given by\cite{wunsch:2006,krstajic:2012}
\begin{equation}
\gamma(q)=\left\vert\frac{{\rm Im}\Pi(q,\omega_p)}{\frac{\partial {\rm Re}\Pi(q,\omega)}{ \partial \omega} \big\vert_{\omega=\omega_p}}\right\vert,
\end{equation}
which we have evaluated numerically.  From this we can see that the upper plasmon branch is indeed undamped for small q within the Pauli blocked region.  However, unlike the case in graphene, where the plasmon branch would encounter the particle-hole continuum and slowly gain a scattering rate\cite{wunsch:2006}, here the branch is deflected and gains a sharp increase in plasmon scattering, which then drops further along the branch.  This behavior is distinct from the plasmons in the linear branch which are always located within the particle-hole continuum.  In this case, the decay rate is always finite and slowly rises as $q$ is increased, exhibiting a broad peak and then decreasing at larger $q$.
Because $\gamma(q)$ rises only slowly with $q$, these excitations, although damped, are still seen to rise above the background in the ${\rm Im}\epsilon^{-1}(q,\omega)$ plot of Fig.~\ref{fig:7}.

It is interesting as to why there are two branches for the inclusion of hexagonal warping, while the pure Dirac case shows only one.  Shown in Fig.~\ref{fig:new}(c) is a comparison of the plasmon poles, which can be expressed as the intersection of $q/(\hbar v \alpha_{FS} ) = (2\pi){\rm Re}\Pi(q,\omega)$.   
The solid black and dashed curves are reproduced from Fig.~\ref{fig:repol}(b) for $\omega=0.5\mu$ and show only their positive part at smaller $q$.  These curves are different from those for the pure Dirac case, which would diverge at $\omega=\hbar v q$.  Here the anisotropy brought in by the hexagonal warping broadens the peak out over a range of momentum $q$ and this leads to two crossings of ${\rm Re}\Pi (q,\omega)$ with the straight line $q/(\hbar v \alpha_{FS})$ and consequently to two plasmon branches.  This is true for both values of $\theta_{\vect{q}}$ shown.  Turning to $\theta_{\vect{q}}=\pi/6$ we show two arrows, black and red that emphasize the plasmon poles for this case and these are further identified in Fig.~\ref{fig:new}(a) along the line $\omega=0.5\mu$ (dashed blue).  As emphasized, the second branch (red squares) will always have a finite lifetime, but our numerical data indicates that it might still be seen in ${\rm Im}\epsilon^{-1}(q,\omega)$ plots.  Returning to Fig.~\ref{fig:new}(c) we can see from this graphical representation that if one reduces $\alpha_{FS}$, then the $q/(\hbar v \alpha_{FS} )$ line will increase in slope.  This will modify the plasmon poles as in Fig.~\ref{fig:wp}(d), and for sufficiently small $\alpha_{FS}$ the $q/(\hbar v \alpha_{FS} )$ line will go above the peak in the real part of the polarization and thus produce no suitable plasmon pole.



\section{Summary and Conclusions}\label{sec:conc}

We have presented numerical results for the real and imaginary parts of the polarization function associated with the surface states of a topological insulator.  The Hamiltonian includes a Dirac term, a sub-dominant Schr\"odinger quadratic, hexagonal warping,  and a gap.  We study how the introduction of these contributions modifies screening and plasmon behavior in these surface states.  In particular, the hexagonal warping introduces an anisotropy into the polarization, $\Pi(\vect{q},\omega)$.  Without this term $\Pi(\vect{q},\omega)$ depends only on the magnitude $|\vect{q}|$ of the momentum transfer but with warping it acquires a dependence on the angle which the scattering momentum, $\vect{q}$, makes with respect to the axis of the 2D surface state Brillouin zone.  This anisotropy is small in the long wavelength limit where $\vect{q}\to 0$, but becomes large as $q$ increases to the order of 0.01~\AA$^{-1}$.  In particular, for a fixed value of $\omega$ the peaks in the imaginary part of the polarization shown in Fig.~\ref{fig:impol} can shift significantly.  Introducing a sub-dominant Schr\"odinger term produces further quantitative changes in $\Pi(\vect{q},\omega)$.  Inclusion of a large gap can result in the splitting of peaks in both the polarization and dielectric functions at large $q$ values.  In all cases an important difference with the pure Dirac case is that the boundaries of the particle-hole continuum shift with inclusion of a warping term which has consequences on plasmon damping processes.  In the pure Dirac case the intraband and interband particle-hole continuum occupy separate parts of the $(q,\omega)$ space.  With warping these regions can overlap.  More importantly for us here the region of no damping becomes smaller and this affects importantly the range of $\omega$ and $q$ for which plasmons remain undamped.  This range also depends on the direction of $\vect{q}$ and is further affected by the introduction of a Schr\"odinger term or of a gap. 

Our numerical calculations have revealed that anisotropy can provided new plasmon branches which, while damped due to falling in the particle-hole continuum, might still be seen in the ${\rm Im}\epsilon^{-1}(q,\omega)$ as peaks above the background because the damping is not large.
Further, we have shown numerically that hexagonal warping restricts the region in $q$ and $\omega$ where the ${\rm Re}\Pi(\vect{q},\omega)>0$ which results in only a small window where plasmons can exist. Finally, we have derived simple analytic formulas for the slope of the plasmon dispersion curve $\omega_p(\vect{q})$ as it comes out of $q=0$.  In all cases it goes  $\propto \sqrt{q}$ but the coefficient of this dependence is changed as warping, Schr\"odinger term and gap are introduced.  Our simple formulas for lowest order corrections confirm our numerical work in that we find the slope towards $q=0$ to be reduced by a gap but increased by warping and Schr\"odinger pieces.  The anisotropy in the plasmon dispersion $\omega_p(\vect{q})$ while in principle always present, is small for small values of $q$ and becomes noticeable only at finite $q$ before the plasmons become damped as the particle hole continuum is reached.  The analytic results we have presented are relevant to recent experimental work\cite{dipietro:2013} which has observed the Dirac plasmons, at very small $q \approx 10^{-5}$~\AA$^{-1}$ in Bi$_2$Se$_3$, while our numerics show deviations from $\sqrt{q}$ behavior at larger scattering momenta which will be relevant to near-field optics techniques which have been successfully applied to graphene.\cite{fei:2012, chen:2012} These results may have further implications aimed at understanding interactions in modified Dirac fermion systems,\cite{carbotte:2012} an area which has yet to be fully explored  in the context of surface states of topological insulators.

\appendix 
\section{Derivation of $q\to 0$ plasmon dispersions}\label{sec:app}
We begin with the energy dispersion for the conduction band for Dirac plus hexagonal warping only.  It has the form
\begin{equation}
E(\vect{k})=\sqrt{ \hbar^2 v^2 k^2 +\lambda^2 \{k_x^3 -3 k_xk_y^2\}^2  }.
\end{equation}
For $E(\vect{k}+\vect{q})$ we define $\theta$ as the angle of $\vect{k}$ and $\alpha$ as the angle of $\vect{q}$ such that the angle between $\vect{k}$ and $\vect{q}$ is $\theta-\alpha$.  In this case, to leading order in the absolute value of $\vect{q}$ the $\lambda^2$ term can be reduced to 
\begin{equation}
\lambda^2 [ k^3 \cos(3\theta) +3 k^2 q \cos(2\theta+\alpha)  ]
\end{equation}
and the energy 
\begin{equation}
E(\vect{k}+\vect{q})=E(k)+q\beta(k,\theta,\alpha)
\end{equation}
where
\begin{equation}\label{eqn:beta}
\beta(k,\theta,\alpha) =  \frac{\hbar^2 v ^2 k \cos(\theta-\alpha)+3\lambda^2 k^5 \cos(3\theta)\cos(2\theta+\alpha)}{E(k)}
\end{equation}
Also the derivative $\left(\frac{d E(k,\theta)}{dk}\right)_{k=k_c(\theta)}$ of Eq.~(\ref{eqn:9}) is given by
\begin{equation}
\left(\frac{d E(k,\theta)}{dk}\right)_{k=k_c(\theta)} = \frac{\hbar^2 v^2 k + 3\lambda^2 k^5 \cos^2(3\theta)}{E(k)} \Bigg|_{k_c(\theta)}
\end{equation}
so that the required integrand of Eq.~(\ref{eqn:9}) is
\begin{align}
I\equiv & \frac{k_c(\theta)\beta^2(k_c(\theta),\theta,\alpha)}{\left(\frac{d E(k,\theta)}{dk}\right)_{k=k_c(\theta)}} \nonumber \\ =&
\frac{k_c [ \hbar^2 v^2 k_c \cos(\theta-\alpha) +3\lambda^2 k_c^5 \cos(3\theta)\cos(2\theta+\alpha)  ]^2}{  E(k_c) [   \hbar^2 v^2 k_c  +3\lambda^2 k_c^5 \cos(3\theta) ]}
\end{align}
where we have suppressed the label $\theta$ on $k_c(\theta)$.  For a given $\theta$ and $\alpha$, $k_c(\theta)$ is given by the solution of $E(k_c(\theta),\theta)=\mu$, which can be simplified to 
\begin{equation}
\hbar^2 v^2 k_c^2 +\lambda^2 k_c^6\cos^2(3\theta)=\mu^2
\end{equation}
which is a cubic equation in $k_c^2$.  We can solve this equation assuming $\lambda$ to be small and then expand to get 
\begin{equation}\label{eqn:a8}
\hbar v k_c \cong \mu \left[1 -\frac{1}{2}\left(\frac{\lambda}{\hbar v}\right)^2 \left(\frac{\mu}{\hbar v}\right)^4 \cos^2(3\theta)\right]
\end{equation}
which leads to 
\begin{equation}\label{eqn:a9}
I\approx
\frac{[ \hbar^2 v^2 k_c^2 \cos(\theta-\alpha) +3 \lambda^2 (\frac{\mu}{\hbar v})^6 \cos(3\theta) \cos(2\theta+\alpha)  ]^2}{\mu[3\mu^2 -2 \hbar^2 v^2 k_c^2]}
\end{equation}
substituting Eq.~\ref{eqn:a8} into the denominator of Eq.~\ref{eqn:a9} we get that \ref{eqn:a9} reduces to
\begin{widetext}
\begin{equation}
I \approx \mu \left[ \cos^2(\theta-\alpha) +\left(\frac{\lambda}{\hbar v}\right)^2 \left(\frac{\mu}{\hbar v}\right)^4  \{ 6\cos(3\theta)\cos(2\theta+\alpha)\cos(\theta-\alpha) -4 \cos^2(3\theta)\cos^2(\theta-\alpha)   \}  \right]
\end{equation}
and after integration over $\theta$ we get
\begin{equation}
I \approx \frac{\mu}{2}\left[1+2 \left(\frac{\lambda}{\hbar v}\right)^2 \left(\frac{\mu}{\hbar v}\right)^4 h(\alpha)\right]
\end{equation}
where $h(\alpha)$ is
\begin{align}
h(\alpha)&=\int\limits_0^{2\pi} \frac{d\theta}{2\pi} [ 6\cos(3\theta) \cos(2\theta+\alpha) \cos(\theta-\alpha) -4 \cos^2(3\theta) \cos^2(\theta-\alpha)   ] \\
 \nonumber &= \frac{1}{2}.
\end{align}
\end{widetext}
We see that  $h(\alpha)$ reduces to $1/2$ and is independent of angle $\alpha$ in this limit, giving
\begin{equation}
\omega_p(q)=\sqrt{ \frac{e^2 \mu}{2\epsilon_0 } q \left[1 + \left(\frac{\lambda}{\hbar v}\right)^2 \left(\frac{\mu}{\hbar v}\right)^4\right] }
\end{equation}

In higher orders however we expect to find anisotropy and the angle $\alpha$ will not drop out as we have established in the numerical work described in the main text.

%
%
%

\begin{acknowledgments}
This research was supported in part by the Natural Sciences and
Engineering Research Council of Canada (NSERC) and the Canadian Institute
for Advanced Research (CIFAR). 
\end{acknowledgments}


\bibliographystyle{apsrev4-1}
\bibliography{bib}

\end{document}